\documentclass[aps,pra,twocolumn,showpacs,preprintnumbers,amsmath,amssymb,footinbib]{revtex4}
\usepackage{graphicx,epsfig}
\usepackage{bm}
\usepackage{dcolumn}
\usepackage{xcolor}
\usepackage[breaklinks=true,colorlinks,citecolor=blue,linkcolor=blue,urlcolor=blue]{hyperref}

\def\nat#1#2#3{Nature {\bf #1}, #2 (#3)}

\def\prv#1#2#3{Phys. Rev. {\bf #1}, #2 (#3)}
\def\rmp#1#2#3{Rev. Mod. Phys. {\bf #1}, #2 (#3)}
\def\prl#1#2#3{Phys. Rev. Lett. {\bf #1}, #2 (#3)}
\def\pra#1#2#3{Phys. Rev. A {\bf #1}, #2 (#3)}
\def\prb#1#2#3{Phys. Rev. B {\bf #1}, #2 (#3)}
\def\prd#1#2#3{Phys. Rev. D {\bf #1}, #2 (#3)}
\def\pre#1#2#3{Phys. Rev. E {\bf #1}, #2 (#3)}
\def\epl#1#2#3{Europhys. Lett. {\bf #1}, #2 (#3)}

\def\epjd#1#2#3{Eur. Phys. J. D {\bf #1}, #2 (#3)}
\def\njp#1#2#3{New J. Phys. {\bf #1}, #2 (#3)}

\def\jpamt#1#2#3{J. Phys. A: Math. Theor. {\bf #1}, #2 (#3)}
\def\jpamg#1#2#3{J. Phys. A: Math. Gen. {\bf #1}, #2 (#3)}
\def\jpb#1#2#3{J. Phys. B: At. Mol. Opt. Phys. {\bf #1}, #2 (#3)}

\def\plb#1#2#3{Phys. Lett. B {\bf #1}, #2 (#3)}

\def\noi{\noindent}
\def\bc{\begin{center}}
\def\ec{\end{center}}
\newcommand{\bea}{\begin{equation}}
\newcommand{\eea}{\end{equation}\noi}
\newcommand{\ber}{\begin{eqnarray}}
\newcommand{\eer}{\end{eqnarray}\noi}
\begin{document}
\title{Casimir effect for a Bose-Einstein condensate inside a cylindrical tube}
\author{Shyamal Biswas$^1$}
\email{sbsp [at] uohyd.ac.in}
\author{Saugata Bhattacharyya$^2$}
\author{Amit Agarwal$^3$}

\affiliation{$^1$School of Physics, University of Hyderabad, C.R. Rao Road, Gachibowli, Hyderabad-500046, India\\
$^2$Department of Physics, Vidyasagar College, 39 Sankar Ghosh Lane, Kolkata-700006, India \\
$^3$Department of Physics, Indian Institute of Technology Kanpur, Kanpur-208016, India}

\date{\today}

\begin{abstract}
We explore Casimir effect on an interacting Bose-Einstein condensate (BEC) inside a cylindrical tube. The Casimir force for the confined BEC comprises of (i) a mean-field part arising from the spatial inhomogeneity of the condensate order parameter, and (ii) a quantum fluctuation part arising from the confinement of Bogoliubov excitations in the condensate. Our analytical result predicts Casimir force on a cylindrical shallow of $^4$He well below the $\lambda$-point, and can be tested experimentally.
\end{abstract}


\maketitle
 
\section{Introduction}
\label{Sec1}

Casimir (like) or fluctuation induced force arises due to confinement of fluctuations, either classical or quantum, and depends upon the confining geometry of the system \cite{Casimir,Krech,Golestanian,Gambassi}. Casimir effect due to the confinement of classical (critical) fluctuations is found to be stronger than the corresponding effect due to the confinement of the quantum (vacuum) fluctuations \cite{Mohideen,Chan2,Hertlein}. Systems with different geometry of the confining plates, eg. sinusoidally corrugated geometry, eccentric-cylindrical geometry, grating geometry, parabolic geometry, concentric-cylindrical geometry, etc, have also been studied both experimentally and theoretically, in the context of the quantum Casimir force \cite{Emig,Chen,Dalvit,Lambrecht,Graham,Teo}. Casimir effects for classical and quantum fluids (specially for superfluid $^4$He) in slab geometry have been the subject of a number of experimental and theoretical works \cite{Krech,Chan2,Biswas2010,Biswas,Dohm2014,Diehl2014}. 
In this context,  Casimir force in a silicon micro-mechanical chip has also been demonstrated, opening up the possibility of tailoring the Casimir force by lithographically made components \cite{chip}. Since the Casimir effect is strongly affected by the confining geometry of the plates, we naturally take up the study of the Casimir force due to the confinement of quantum fluctuations in a cylindrical geometry.  

Casimir effect due to confinement of quantum (vacuum) fluctuations of electromagnetic field in a hollow cylindrical tube was studied initially by DeRaad Jr. and Milton in Ref.~\cite{Milton1981}, and later on by a number of authors \cite{Gosdzinsky,Milton1999,Milton2010,Straley,Milton2004}. In this article, we consider a Bose-Einstein condensate confined to the cylindrical geometry, and explore the Casimir force arising from the presence of the condensate inside the cylinder. For a confined BEC, while the Bogoliubov excitations in the condensate are responsible for the quantum fluctuation part of the Casimir force, the confinement of the condensate wave function, which obeys Gross-Pitaevskii equation, gives rise to the mean-field part of Casimir force. We explicitly compute these two contributions to the Casimir effect.

We begin this article by revisiting the calculation of the Casimir force due to the confinement of electromagnetic field in a hollow conducting circular-cylinder. Bessel functions [$J_\sigma(\alpha_{\sigma,n}\rho/R)$ and $J_\sigma(\beta_{\sigma,n}\rho/R)$], Bessel zeros ($\alpha_{\sigma,n};~\sigma=0,\pm1,\pm2,...;~n=1,2,3,...$), and zeros of the first derivatives of the Bessel functions of the first kind ($\beta_{\sigma,n}$) naturally appear in the modes of quantum fluctuations and in the expression of the vacuum energy. To get the Casimir force from the vacuum energy we apply zeta regularization technique prescribed by Gosdzinsky and Romeo \cite{Gosdzinsky}. Then by applying these techniques, we study the Casimir effect filling the hollow cylinder by an interacting Bose-Einstein condensate for $T \to 0$. For this case, we begin with the calculation of the grand canonical energy of the interacting BEC in terms of its mean-field and quantum fluctuation (discrete Bogoliubov excitation) parts. We explicitly 
calculate the contribution to the Casimir force arising from the mean-field and quantum fluctuations parts for both the Dirichlet and the Neumann boundary conditions. Thus we study condensate density dependence of the Casimir force. We apply dimensional regularization cum Chowla-Selberg lattice summation technique \cite{Milton2010} for the higher orders in condensate density. Finally, we summarize our results in the conclusion.

\section{Casimir effect on the hollow cylinder}
\label{Sec2}
\begin{figure}
\includegraphics[width=.95 \linewidth]{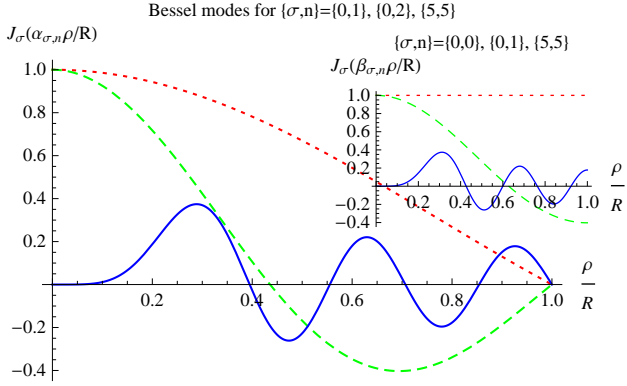}
\caption{(Color online) Bessel modes for Dirichlet and Neumann (inset) boundary conditions. 
\label{fig1}}
\end{figure}

Before discussing the Casimir effect for a Bose-Einstein condensate enclosed in a circular-cylinder geometry, let us review the conventional Casimir effect on a hollow circular-cylinder. This will also enable us to establish the condensate-density dependence on the Casimir force, and the validity of our approximations later. Let us consider an infinitely long hollow circular-cylinder of radius $R$ and length $L$, whose axis is along the $z$-direction, and whose radial and angular coordinates are represented by $(\rho, \theta)$. The zero point or vacuum energy of the electromagnetic field inside the cylinder is given by \cite{Casimir}
\begin{eqnarray}\label{eqn:11}
E=\frac{1}{2}\sum_{\bm{k}}\hbar\omega_{\bm{k}}~,
\end{eqnarray}
$\bm{k}$ represents all possible (TM and TE) modes of the vacuum fluctuations compatible to the boundary conditions imposed by the waveguide \textit{i.e.} the circular-cylindrical conductor. While in free space or in the space between two parallel conductors, the two polarization states of the electromagnetic field are equivalent; in the space inside a cylindrical waveguide, the two polarization states known as transverse magnetic (TM) and transverse electric (TE) modes are not equivalent. For the conducting cylindrical shell surface, the TM modes of electromagnetic fluctuations ($B_z=0$ and $E_z|_{\rho=R}=0$) which obey Dirichlet boundary condition are given by the Bessel functions (so that $E_z=J_\sigma(\alpha_{\sigma,n}\rho/R)\text{e}^{i\sigma\theta}\text{e}^{ikz}$), and the Bessel zeros specified by $\alpha_{\sigma,n}$, with  $\sigma=0,\pm1,\pm2,...$ and $n=1,2,3,...$, naturally appear in the expression of the vacuum energy \cite{Jackson}. On the other hand, the TE modes ($E_z=0$ and $\frac{\partial B_z}{
\partial\rho}|_{\rho=R}=0$) inside the cylinder are compatible with the Neumann boundary condition, and they are given by $B_z=J_\sigma(\beta_{\sigma,n}\rho/R)\text{e}^{i\sigma\theta}\text{e}^{ikz}$ where $\beta_{\sigma,n}$s are positive zeros of the first derivatives of the Bessel functions of the first kind \cite{Jackson}. The two distinct class of modes, as shown in Fig.~\ref{fig1}, contribute simultaneously to the vacuum energy of the conducting cylinder, which can be obtained from Eq.~(\ref{eqn:11}) to be \cite{Gosdzinsky,Milton1999,Jackson}
\begin{eqnarray}\label{eqn:12}
E= \frac{\hbar c L}{2}\sum_{\sigma,n} \int_{-\infty}^\infty \frac{dk}{2\pi}\bigg[\sqrt{\frac{\alpha_{\sigma,n}^2}{R^2}+k^2}+\sqrt{\frac{\beta_{\sigma,n}^2}{R^2}+k^2}\bigg],
\end{eqnarray}
where the first and second terms in the square bracket represent contributions for the TM modes and TE modes respectively. Note that the integral over $k$ in the above expression has a logarithmic divergence which may not be easily removed by applying standard analytic continuation techniques. What work here are ultraviolet cutoff technique with Green’s dyadic formulation for the field strengths \cite{Milton1981}, zeta regulation technique \cite{Gosdzinsky}, mode-by-mode summation technique \cite{Milton1999}, and, to some extent, dimensional regularization technique with Chowla-Selberg lattice summation \cite{Milton2010}.

Integrals in Eq.~\eqref{eqn:12} can be evaluated within dimensional regularization scheme. The first integral can be recast with the substitution $q^2=\frac{\alpha_{\sigma,n}^2}{R^2}+k^2$, as $I_1=2\int_{\alpha_{\sigma,n/R}}^\infty \frac{dq}{2\pi}q^2(q^2-\frac{\alpha_{\sigma,n}^2}{R^2})^{-1/2}$, which for the sake of dimensional regularization, can be further recast as $I_1=2\int_{\alpha_{\sigma,n/R}}^\infty \frac{dq}{2\pi}q^{-2s}(q^2-\frac{\alpha_{\sigma,n}^2}{R^2})^{-d}$ where $s=-1$ and $d=1/2$. It is easy to check that, this integral diverges for $s=-1$ and $d=1/2$. This divergence can be avoided within dimensional regularization scheme by substituting $d=1/2+\epsilon$ (where $\epsilon\rightarrow0$) and $s=-1+\delta$ (where $\delta\rightarrow0$), and by subsequently performing analytic continuation of gamma and zeta functions for $\Re[s]+d\le1/2$. In a similar way, we can regularize the second integral in Eq.~\eqref{eqn:12}. Thus within the dimensional regularization scheme, Eq.~\eqref{eqn:12} can be recast as
\begin{eqnarray}\label{eqn:13o0}
E=L\hbar c[D^{(-2s)}+N^{(-2s)}], 
\end{eqnarray}
where keeping $s\rightarrow-1$ unaltered we put $d=1/2$ after evaluating both the integrals in Eq.~\eqref{eqn:12}, as,
\begin{eqnarray}\label{eqn:13o1}
D^{(-2s)}=\frac{1}{4\sqrt{\pi}}\sum_{\sigma=-\infty,n=1}^{\infty,\infty}\left(\frac{\alpha^2_{\sigma,n}}{R^2}\right)^{-s}\frac{\Gamma(s)}{\Gamma(\frac{1+2s}{2})},
\end{eqnarray}
and
\begin{eqnarray}\label{eqn:13o2}
N^{(-2s)}=\frac{1}{4\sqrt{\pi}}\sum_{\sigma=-\infty,n=0}^{\infty,\infty}\left(\frac{\beta^2_{\sigma,n}}{R^2}\right)^{-s}\frac{\Gamma(s)}{\Gamma(\frac{1+2s}{2})}.
\end{eqnarray}
Here `D' and `N' respectively stand for Dirichlet and Neumann boundary conditions. Regularization of $D^{(-2s)}$ in Eq.~\eqref{eqn:13o1} and that of $N^{(-2s)}$ in Eq.~\eqref{eqn:13o2}, with the use of Chowla-Selberg lattice-sum formula \cite{Chowla-Selberg}, needs exact values of $\alpha_{\sigma,n}$ and $\beta_{\sigma,n}$ which although appear in closed forms for triangular-cylinder, square-cylinder, \textit{etc}, unfortunately do not appear in closed forms for circular-cylinder. For more details regarding the Chowla-Selberg lattice summation, see-- Appendix-A of Ref.~\cite{Milton2010}, and also our Appendix-\ref{A}. Another problem with the Chowla-Selberg formula for circular-cylinder is that, it is not even applicable for the McMahon asymptotic forms of $\alpha_{\sigma,n}\rightarrow\pi(\sigma/2+n-1/4)$ and $\beta_{\sigma,n}\rightarrow\pi(\sigma/2+n+1/4)$ \cite{McMahon} as the `determinant'\footnote{If $\alpha^2_{\sigma,b}=\pi^2(a\sigma^2+b\sigma n+cn^2)$, then the `determinant' is defined as $\
triangle=4ac-b^2$.} $\triangle\rightarrow0$. Exact renormalized values of $D^{(2)}$ and $N^{(2)}$ were obtained, within the integral forms of the lattice summations (which do not at all need the values of $\alpha_{\sigma,n}$ and $\beta_{\sigma,n}$) with further subtraction of homogeneous energy eigenvalue $\omega_{\bm k}|_{R\rightarrow\infty}$ from each energy eigenvalue $\omega_{\bm k}$ in Eq.~\eqref{eqn:11}, as \cite{Milton1999,Gosdzinsky}
\begin{eqnarray}\label{eqn:13o3}
D^{(2)}=0.000614794033/R^2=0.0019314324176/\pi R^2,
\end{eqnarray}
and 
\begin{eqnarray}\label{eqn:13o4}
N^{(2)}=-0.01417613719/R^2=-0.0445356485/\pi R^2.
\end{eqnarray}
Thus we get the total Casimir (or renormalized vacuum) energy as \cite{Milton1981,Gosdzinsky}
\begin{eqnarray}\label{eqn:18}
E_c=L\hbar c[D^{(2)}+N^{(2)}]=-0.013561343{\small{~}}L\hbar c/R^2,
\end{eqnarray}
and the Casimir force ($f_R=-\frac{\partial E_c}{\partial R}$) as 
\begin{eqnarray}\label{eqn:19a}
f_R=-0.027122686{\small{~}}L\hbar c/R^2.
\end{eqnarray}
The Casimir pressure ($p_R=f_R/2\pi RL$) is now given by $p_R\approx-0.0043\frac{\hbar c}{R^4}$, which is reasonably strong as $R$ approaches the atomic dimension. For example, if $R=10~\text{\AA}$, the Casimir pressure is $p_R\approx-1\times10^{8}$ N/m$^2$. Such a strong attractive Casimir pressure can possibly be tested for cylindrical graphene sheet, carbon nano-tube, etc. For a circular waveguide of radius $R=0.1$~mm, on the other hand, the Casimir pressure is $p_R\approx-1\times10^{-12}$ N/m$^2$. It is interesting to note that confinement of TE modes dominates that of TM modes, and that $N^{(2)}$ per cross-sectional area of a square-cylinder ($-0.0429968$ \cite{Milton2010}) is close to that of the circular-cylinder in Eq.~\eqref{eqn:13o4} with deviation less than $4\%$. This brings a hope that the Casimir effect per cross-sectional area of circular-cylinder would be close to that of a square-cylinder as both are topologically same, and as the later specially is having 4-fold symmetry. 

\section{Casimir effect for a BEC inside the circular-cylinder}
\label{Sec3} 
The Casimir force on an interacting Bose-Einstein condensate was studied by Biswas \textit{et al} in Ref.~[\onlinecite{Biswas2010}] for a film geometry. Here we generalize the same (in a nontrivial manner) for an interacting BEC confined to a cylindrical geometry. 

\subsection{Elementary excitations in the BEC inside the cylindrical tube}
Let us consider a Bose gas of $N$ identical particles be kept inside the same cylinder of radius $R$ and length $L$, and take the same set of cylindrical coordinates ($\rho,\theta,z$) to specify position $\textbf{r}$ as in the previous section. For $T\rightarrow 0$, almost all the particles ($N_0\approx N$) form the condensate, and the elementary (phononic as well as Bogoliubov) excitations in the condensate can be treated perturbatively. The elementary excitations of a BEC are generally described by expanding the Hamiltonian up-to the second order in quantum fluctuations over the BEC order parameter ($\sqrt{N_0}\phi_0(\bf r)$), via  a standard perturbative approach  \cite{Biswas2010}. We follow a similar approach here for obtaining the 
elementary excitations for the cylindrical geometry. 

Let the position vector and mass of a single Bose particle be denoted as ${\bf r}$ and $M$ respectively. The grand canonical Hamiltonian operator for such a system of interacting Bose gas is given by \cite{Biswas2010}
\begin{eqnarray}\label{eqn:61}
\hat{\mathcal H}&=&\int\hat{\Psi}^{\dagger}({\bf r})\bigg(-\frac{\hbar^2}{2M}\nabla^2-\mu\bigg)\hat{\Psi}({\bf r})\text{d}^3{\bf r}\nonumber \\ &+&\frac{1}{2}\int\int\bigg(\hat{\Psi}^{\dagger}({\bf r})\hat{\Psi}^{\dagger}({\bf r'})V({\bf r-r'})\hat{\Psi}({\bf r})\hat{\Psi}({\bf r'})\bigg)\text{d}^3{\bf r}\text{d}^3{\bf r'},~~~
\end{eqnarray}
where $\mu$ is the chemical potential, $V({\bf r-r'})$ is the inter-particle interaction potential, and $\hat{\Psi}({\bf r})$ is the field operator for Bose particles. For the simplest case, the interaction potential can be considered as $V({\bf r-r'})=g\delta^3({\bf r-r'})$, where $g=4\pi\hbar^2a_s/M$ is the coupling constant and $a_s$ is the s-wave scattering length \cite{Biswas2010}. The field operator can be expressed in terms of the single particle orthonormal wave functions $\{\phi_i({\bf r})\}$,  as $\hat{\Psi}({\bf r})=\sum_{i=0}^{\infty}\phi_i({\bf r})\hat{a}_i$, where $\hat{a}_i$ ($\hat{a}_i^{\dagger}$) annihilates (creates) a Bose particle in the state $\phi_i({\bf r})$. Within a perturbative approach, the grand canonical Hamiltonian can be expressed in terms of the fluctuations (excitations) $\delta\hat{\Psi}({\bf r})=\hat{\Psi}({\bf r})-\sqrt{N_0}\phi_0({\bf r})$, and up-to quadratic order in $\delta \hat{\Psi}$, it is given by \cite{Biswas2010}
\begin{eqnarray}\label{eqn:62}
\hat{\mathcal H}&=&\Omega_0+\int d^3{\bf r} \left[\delta\hat{\Psi}^{\dagger}({\bf r})\bigg(-\frac{\hbar^2}{2M}\nabla^2\bigg)\delta\hat{\Psi}({\bf r})\text{d}^3{\bf r} + \frac{g\bar{n}}{2} \right. \nonumber \\ 
&\times&\left. \bigg(2\delta\hat{\Psi}^{\dagger}({\bf r})\delta\hat{\Psi}({\bf r})+\delta\hat{\Psi}^{\dagger}({\bf r})\delta\hat{\Psi}^{\dagger}({\bf r})
+ \delta\hat{\Psi}({\bf r})\delta\hat{\Psi}({\bf r})\bigg) \right],~~~
\end{eqnarray}
where $\bar{n}\approx N_0\phi_0^2(\textbf{r})$ denotes the bulk particle density, $\mu\approx g\bar{n}$, and $\Omega_0$ is the  grand potential for the condensate, which is given by \cite{Biswas2010}
\begin{eqnarray}\label{eqn:63}
\frac{\Omega_0}{N_0} = \int\bigg(\frac{\hbar^2}{2M}|\nabla\phi_0(\textbf{r})|^2-\mu|\phi_0(\textbf{r})|^2+\frac{gN_0}{2}|\phi_0(\textbf{r})|^4\bigg)\text{d}^3{\textbf{r}}.
\end{eqnarray}
In a Bose-Einstein condensate low energetic excitations are probabilistically favored as temperature of the system decreases. We already have experienced in the previous section that, while the Casimir energy for the TM modes (which obey Dirichlet boundary condition) are positive, that for the TE modes (which obey Neumann boundary condition) are negative. Thus, the natural boundary condition for the elementary excitations in the condensate in an infinitely long cylinder, is expected to be of Neumann type. However, by the definition of the quantum vacuum energy, contribution of the elementary excitations of Dirichlet type is also unavoidable even for $T\rightarrow0$ unless any filtering of modes is engineered (e.g. like that of a resonant cavity) at the two ends of the cylinder. The quantum fluctuations (\textit{i.e.} the elementary excitations) for the Neumann boundary on a cylindrical tube can be expressed as $\delta\hat{\Psi}(\textbf{r})= L (2 \pi \hbar)^{-1} \sum_{\sigma=-\infty,
n=1}^{\infty}\int_{-\infty}^\infty \phi_{p_z,\sigma,n} \hat{a}_{p_z,\sigma,n} {d}p_z$, where $\hat{a}_{p_z,\sigma,n}$ annihilates a Bose particle specified by the state $\phi_{p_z,\sigma,n}(z,\rho,\theta)=\sqrt{\frac{2}{R^2L}}\text{e}^{i\frac{p_zz}{\hbar}}\frac{e^{i\sigma\theta}}{\sqrt{2\pi}}\frac{J_\sigma(\beta_{\sigma,n}\rho/R)}{J_{\sigma+1}(\beta_{\sigma,n})}$ \cite{Cavity-Modes}, whose momentum in  the $z$-direction is $p_z$ and whose energy is  $\xi=\frac{p_z^2}{2M}+\frac{\hbar^2\beta_{\sigma,n}^2}{2MR^2}$. The Hamiltonian in Eq.~\eqref{eqn:62} can now be diagonalized in terms of the phononic operators through the usual Bogoliubov transformation: \cite{Roberts} $\hat{a}_{p_z,\sigma,n}=u_{p_z,\sigma,n}\hat{b}_{p_z,\sigma,n}+v_{p_z,\sigma,n}\hat{b}_{-p_z,\sigma,n}^{\dagger}$~,  
where $u_{p_z,\sigma,n}=\big(\frac{\xi^2+g\bar{n}}{2\epsilon_N(|p_z|,\sigma,n)}+\frac{1}{2}\big)^{1/2}$, $v_{p_z,\sigma,n}=-\big(\frac{\xi^2+g\bar{n}}{2\epsilon_N(|p_z|,\sigma,n)}-\frac{1}{2}\big)^{1/2}$, and
\begin{equation}\label{eqn:64}
\epsilon_N(|p_z|,\sigma,n)=\bigg[\frac{g\bar{n}}{M}\bigg(p_z^2+\frac{\hbar^2\beta_{\sigma,n}^2}{R^2}\bigg)\bigg(1+\frac{p_z^2+\frac{\hbar^2\beta_{\sigma,n}^2}{R^2}}{4Mg\bar{n}}\bigg)\bigg]^{\frac{1}{2}}. \\
\end{equation}

For the TM modes (i.e. for the Dirichlet boundary condition), on the other hand, the Bogoliubov excitation energy would be $\epsilon_D(|p_z|,\sigma,n)$ which is almost like $\epsilon_N(|p_z|,\sigma,n)$ in Eq.~\eqref{eqn:64} except $\beta_{\sigma,n}$ be replaced by $\alpha_{\sigma,n}$ everywhere. For the TM modes, we also call the phononic annihilation and creation operators by $\hat{d}_{p_z,\sigma,n}$ and $\hat{d}_{p_z,\sigma,n}^{\dagger}$ respectively instead of $\hat{b}_{p_z,\sigma,n}$ and $\hat{b}_{p_z,\sigma,n}^{\dagger}$ used for the TE modes.

Thus from Eqs.~\eqref{eqn:62} and \eqref{eqn:64}, and also from the subsequent discussion about the TM modes, we have the diagonalized Hamiltonian in terms of both the (Neumann and Dirichlet) types of the Bogoliubov excitations, as 
\begin{eqnarray}\label{eqn:65}
\hat{{\mathcal H}}&=& \Omega_0+{\mathcal E}_0(R,\bar{n}) + \sum_{p_z,\sigma,n}\epsilon_N(|p_z|,\sigma,n)\hat{b}_{p_z,\sigma,n}^{\dagger}\hat{b}_{p_z,\sigma,n}\nonumber\\&&+ \sum_{p_z,\sigma,n}\epsilon_D(|p_z|,\sigma,n)\hat{d}_{p_z,\sigma,n}^{\dagger}\hat{d}_{p_z,\sigma,n}, 
\end{eqnarray}
where 
\begin{eqnarray}
{\mathcal E}_0(R,\bar{n})&=&\frac{1}{2}\sum_{p_z,\sigma,n}\bigg[ \epsilon_N(|p_z|,\sigma,n)-\bigg(\frac{p_z^2}{2M}+\frac{\hbar^2\beta_{\sigma,n}^2}{2MR^2}\bigg)-g\bar{n}\bigg]\nonumber\\&+&\frac{1}{2}\sum_{p_z,\sigma,n}\bigg[\epsilon_D(|p_z|,\sigma,n)-\bigg(\frac{p_z^2}{2M}+\frac{\hbar^2\alpha_{\sigma,n}^2}{2MR^2}\bigg)-g\bar{n}\bigg]\nonumber\\
\end{eqnarray} 
is the contribution to the grand canonical potential due to the quantum  fluctuations of the phonon field. For $T\rightarrow 0$, there are no phonons, and consequently, the Hamiltonian of the system for the vacuum of the phonons is specified by $ {\mathcal H}=\Omega_0+{\mathcal E}_0(R,\bar{n}) $. Roberts and Pomeau  also deduced a similar term for a homogeneous condensate in Ref.~\cite{Roberts} based on the original work of Lee, Huang and Yang \cite{Lee}.

\subsection{Quantum fluctuation part of the Casimir force}
In this subsection we calculate the contribution to the Casimir force arising from the second term in Eq.~\eqref{eqn:65}, {\it i.e.~} the quantum (vacuum) fluctuation part (${\mathcal E}_0$) of the grand canonical Hamiltonian ($\hat{{\mathcal H}}$). Before proceeding with further calculations, we anticipate that the Casimir force due to vacuum fluctuations of the phonon field ($\delta\hat{\Psi}$) would be similar to that due to the photon (electromagnetic) field in Eq.~\eqref{eqn:19a} with the replacement of the speed of light by the speed of phonon (sound), $c \to v(\bar{n})=\sqrt{g\bar{n}/M}$. Thus, using Eqs.~\eqref{eqn:13o3}, \eqref{eqn:13o4}, and \eqref{eqn:19a}, we expect the phonon Casimir force to be 
\begin{eqnarray}\label{eqn:67}
f_R^{(\text{phn})}=\frac{2L\hbar v(\bar{n})}{R}[N^{(2)}+D^{(2)}]\approx-0.027122686\frac{\hbar v(\bar{n})L}{R^3}.~~
\end{eqnarray}

Let us now go back to the Eq.~\eqref{eqn:65} whose second term ${\mathcal E}_0(R,\bar{n})$, gives the quantum vacuum fluctuations as well as the quantum Casimir force for phonons. Expanding $\epsilon_N(|p_z|,\sigma,n)$ in terms of $p=\sqrt{p_z^2+\frac{\hbar^2\beta_{\sigma,n}^2}{R^2}}$ and $\epsilon_D(|p_z|,\sigma,n)$ in terms of $p'=\sqrt{p_z^2+\frac{\hbar^2\alpha_{\sigma,n}^2}{R^2}}$, we have
\begin{eqnarray}\label{eqn:68}
{\mathcal E}_0(R,\bar{n})&=&\frac{1}{2}\sum_{p_z,\sigma,n}\bigg(v(\bar{n})p\left[1-\frac{p}{\sqrt{4Mg\bar{n}}}+\frac{p^2}{8Mg\bar{n}} \right.\nonumber\\
&-& \left. \frac{p^4}{128M^2g^2\bar{n}^2}+...\right]-g\bar{n}\bigg)\nonumber\\&&+\frac{1}{2}\sum_{p_z,\sigma,n}\bigg(v(\bar{n})p'\left[1-\frac{p'}{\sqrt{4Mg\bar{n}}}+\frac{p'^2}{8Mg\bar{n}} \right.\nonumber\\
&-& \left. \frac{p'^4}{128M^2g^2\bar{n}^2}+...\right]-g\bar{n}\bigg).
\end{eqnarray}
Here the first and the second summations correspond to the TE and TM modes respectively. Now, replacing the sum over $p_z$ by an integration over $p_z$, we can rewrite Eq.~\eqref{eqn:68} as
\begin{eqnarray}\label{eqn:69}
{\mathcal E}_0(R,\bar{n})&=&L\bigg(v(\bar{n})\bigg[\hbar (N^{(2)}+D^{(2)})-\frac{\hbar^{2}(N^{(3)}+D^{(3)})}{\sqrt{4Mg\bar{n}}}\nonumber\\
&&+\frac{\hbar^{3}(N^{(4)}+D^{(4)})}{8Mg\bar{n}}-\frac{\hbar^{5}(N^{(6)}+D^{(6)})}{128M^2g^2\bar{n}^2}+...\bigg]\nonumber\\
&&-g\bar{n}(N^{(1)}+D^{(1)})\bigg).
\end{eqnarray}
where $q=p/\hbar$, $N^{(-2s)}=\frac{1}{2\pi}\sum_{\sigma,n}\int_{\frac{\beta_{\sigma,n}}{R}}^\infty q^{-2s}(q^2-\beta_{\sigma,n}^2/R^2)^{-d}= \frac{1}{4\sqrt{\pi}}\sum_{\sigma,n}\left(\frac{\beta^2_{\sigma,n}}{R^2}\right)^{-s}\frac{\Gamma(s)}{\Gamma(\frac{1+2s}{2})}$, $q'=p'/\hbar$, $D^{(-2s)}=\frac{1}{2\pi}\sum_{\sigma,n}\int_{\frac{\alpha_{\sigma,n}}{R}}^\infty q'^{-2s}(q'^2-\alpha_{\sigma,n}^2/R^2)^{-d}= \frac{1}{4\sqrt{\pi}}\sum_{\sigma,n}\left(\frac{\beta^2_{\sigma,n}}{R^2}\right)^{-s}\frac{\Gamma(s)}{\Gamma(\frac{1+2s}{2})}$, $d\rightarrow1/2$, and $s\rightarrow-1/2,-1,-3/2,-2,-5/2,...$. We already have mentioned various regularization method for $N^{(-2s)}$ and $D^{(-2s)}$ in Sec.~\ref{Sec2}. Among such integrals cum lattice summations, $N^{(2)}$ and $D^{(2)}$ can be exactly obtained within zeta regularization scheme as we have already mentioned: $N^{(2)}=-0.01417613719/R^2$ and $D^{(2)}=0.000614794033/R^2$ \cite{Gosdzinsky}. It is absolutely not easy to obtain regularized values of $N^{(-2s)}$ and $N^{(-2s)}$ for $s\le0$ generalizing the works of the authors of Refs. \cite{Gosdzinsky} and \cite{Milton1999}. At the end of Sec.~\ref{Sec2}, we have argued that the Casimir force per cross-sectional area of circular-cylinder would be close to that of a square-cylinder as both are topologically same, and as the later specially is having 4-fold symmetry. 
Evaluation of $N^{(-2s)}$ is reasonably easy for square-cylinder within dimensional regularization and subsequent Chowla-Selberg lattice summation technique \cite{Milton2010}. See-- Appendix-\ref{A} for the use of Chowla-Selberg lattice summation in this subsection too. Following this technique we get $N^{(4)}=0.00914223/R^4$, $D^{(4)}=-0.000263472/R^4$, $N^{(6)}=-0.0185855/R^6$, $D^{(6)}=-0.000389195/R^6$, $N^{(8)}=0.0745135/R^8$, $D^{(8)}=0.00400674/R^8$, and $N^{(1)}=D^{(1)}=N^{(3)}=N^{(D)}=N^{(5)}=D^{(5)}=0$. Substituting these values of $N^{(-2s)}$ and $D^{(-2s)}$ in Eq.~\eqref{eqn:69}, we have 
\begin{eqnarray}\label{eqn:70}
{\mathcal E}_0(R,\bar{n})&=&-\frac{0.013561343~\hbar v(\bar{n})L}{R^2}\bigg(1\nonumber\\
&-&\frac{0.008878758}{0.013561343\times4\tilde{\rho}(R,g\bar{n})} \\
&-&\frac{0.018974695}{0.013561343\times32\tilde{\rho}^2(R,g\bar{n})}+...\bigg),~ \nonumber 
\end{eqnarray}
where $\tilde{\rho}(R,g\bar{n})=\frac{g\bar{n}2MR^2}{\hbar^2}$ is the dimensionless density. The quantum (vacuum) fluctuation part of the Casimir force, $f_R^{\text{(qf)}}=-\frac{\partial}{\partial R}{\mathcal E}_0(R,\bar{n})$, is now easily obtained from Eq.~\eqref{eqn:70}, as
\begin{eqnarray}\label{eqn:71}
f_R^{(\text{qf})}=-0.019\frac{\hbar^2L}{MR^4}\tilde{\rho}^{1/2}\bigg(1-\frac{0.655}{\tilde{\rho}}-\frac{0.262}{\tilde{\rho}^2}+...\bigg).~~
\end{eqnarray}
Note that the first term of the Eq.~\eqref{eqn:71} is the most dominating term for the quantum phonon fluctuations part of the Casimir force, and it is identical to Eq.~\eqref{eqn:67}, which is just based on 
an intuitive similarity to the case of Casimir force arising the TE modes of the vacuum fluctuations of the electromagnetic field.

\subsection{Mean field part of the Casimir force}
In this subsection we calculate the contribution to the Casimir force arising from the first term in Eq.~\eqref{eqn:65}, {\it i.e.~} the mean field part ($\Omega_0$) of the grand canonical Hamiltonian (${\mathcal H}$). Mean-field part of the Casimir force arises due to inhomogeneity of the BEC order parameter [$m(\textbf{r})=\sqrt{N_0}\phi_0(\textbf{r})$] in $\Omega_0$. If the order parameter is complex, it can be expressed as $m(\textbf{r})=m(\rho)\text{e}^{i(\sigma\theta+kz)}$ where the amplitude $m(\rho)$ takes care of inhomogeneity of the condensate density ($n_0(\textbf{r})=|m(\textbf{r})|^2$) along the radial direction and the phase  $\sigma\theta+kz$ accounts for the flow of the condensate either rotationally along the $\hat{\theta}$ direction or translationally along the $\hat{z}$ direction. For simplicity, we consider the condensate to be static inside the cylinder so that the condensate order parameter can be written as a real function of the radial 
coordinate as $m(\textbf{r})=m(\rho)$. We can express the Ginzburg-Landau form in Eq.~\eqref{eqn:mf1} in terms of the BEC order parameter $m(\rho)$ as,
\begin{eqnarray}\label{eqn:mf1}
\Omega_0=\frac{\hbar^2N_0}{M}\int\bigg[\frac{1}{2}[\nabla m(\rho)]^2-\frac{|a|}{2}m^2(\rho)+\frac{b}{4}m^4(\rho)\bigg]\text{d}^3\textbf{r},~
\end{eqnarray}
where we have defined, ${a} = \frac{2M\mu}{\hbar^2}=\frac{2Mg\bar{n}}{\hbar^2}$  and $ b = \frac{2MgN_0}{\hbar^2}>0$.
The minimization of the energy in Eq.~\eqref{eqn:mf1} with respect to the order parameter $m(\rho)$, {\it i.e.~}$ \frac{\delta\Omega_0}{\delta m}=0$, leads to the following equation for the profile of the order parameter,
\begin{equation}\label{eqn:39}
-\nabla^2m-|a|m+bm^3=0~,
\end{equation}
which can be rewritten in terms of the cylindrical polar coordinates as 
\begin{equation}\label{eqn:40}
-\frac{\text{d}^2m}{\text{d}\rho^2}-\frac{1}{\rho}\frac{\text{d}m}{\text{d}\rho}-|a|m+bm^3=0~.
\end{equation}
Note that, for a large system, the order parameter is expected to be homogeneous, and in the bulk limit we have $m \to m_0=\sqrt{|a|/b}$. 

The mean-field component of the force is given by $F_R^{(\text{mf})}=-\frac{\delta \Omega_0}{\delta R}$. It is not necessary that the boundary condition on the condensate has to be the same as that assigned to the elementary excitation in the condensate as they are independent. In a realistic situation of a superfluid/condensate in contact with a metallic plate the boundary condition is considered to be of Dirichlet type \cite{Chan,Krech,Biswas}. For Dirichlet boundary condition on a cylindrical surface and the ground sate of the system, {\it i.e.~} for $ m(R)=0, \frac{dm}{d\rho}|_{\rho=0}=0$\footnote{For the ground state, $m(\rho)=\sum_{n=1}^\infty a_n\cos(\frac{n\pi r}{2R})$, which leads to $\frac{\text{d}m}{\text{d}\rho}|_{\rho=R}=-\frac{dm}{dR}$.}, 
we obtain, 
\begin{eqnarray}\label{eqn:42}
F_R^{(\text{mf})}=\frac{\pi\hbar^2N_0}{M} LR\bigg(\frac{\text{d}m}{\text{d}\rho}\bigg)^2\bigg|_{\rho=R}.
\end{eqnarray}
Eq.~\eqref{eqn:42} is physically appealing as it relates the energy density at the boundary ($\propto (d_\rho m|_R)^2$) to the mean-field part of the Casimir pressure, and a similar form is also expected in other confined geometries \cite{Biswas2010}. The value of $\big(\frac{\text{d}m}{\text{d}\rho}\big)\big|_{\rho=R}$ is to be obtained from the non-linear Eq.~\eqref{eqn:40} which does not have a closed form analytical solution to the best of our knowledge. However, we can obtain an approximate form for the same. Multiplying Eq.~\eqref{eqn:40} by $2\frac{\text{d}m}{\text{d}\rho}$ and integrating over $\rho$, we get
\begin{eqnarray}\label{eqn:43}
\left(\frac{\text{d}m}{\text{d}\rho}\right)^2+\int_0^\rho\frac{2}{\rho}\left(\frac{\text{d}m}{\text{d}\rho}\right)^2\text{d}\rho&=&-|a|\left[m^2-m^2(0)\right]\nonumber\\&+&\frac{b}{2}\left[m^4-m^4(0)\right].
\end{eqnarray}
The second term in the LHS in Eq.~\eqref{eqn:43}, can be integrated by parts:  $\int_0^\rho\frac{2}{\rho}\big(\frac{\text{d}m}{\text{d}\rho}\big)^2\text{d}\rho=2\big(\frac{\text{d}m}{\text{d}\rho}\big)^2|_{0}^\rho-\{\frac{\text{d}}{\text{d}\rho}\big(\frac{1}{\sqrt{\rho}}\frac{\text{d}m}{\text{d}\rho}\big)^2\}\rho^2|_{0}^\rho+\int_0^\rho\{\frac{\text{d}^2}{\text{d}\rho^2}\big(\frac{1}{\sqrt{\rho}}\frac{\text{d}m}{\text{d}\rho}\big)^2\}\rho^2\text{d}\rho$.  Now as $\rho\rightarrow0$, $\frac{1}{\sqrt{\rho}}\frac{\text{d}m}{\text{d}\rho}\rightarrow0$ for the ground state. Additionally, the derivatives of $\frac{1}{\sqrt{\rho}}\frac{\text{d}m}{\text{d}\rho}$ can also be neglected in comparison to the first term as the order parameter for the ground state is smoothest.  Under these approximations, $\int_0^\rho\frac{2}{\rho}\big(\frac{\text{d}m}{\text{d}\rho}\big)^2\text{d}\rho \approx 2\big(\frac{\text{d}m}{\text{d}\rho}\big)^2$.
Thus  Eq.~(\ref{eqn:43}) can be rewritten as $-\int_{m(0)}^0\frac{\text{d}m}{\sqrt{-|a|\big(m^2-m^2(0)\big)+\frac{b}{2}\big(m^4-m^4(0)\big)}}\approx\int_0^R\frac{\text{d}\rho}{\sqrt{3}}$,  which in turn can be expressed as
\begin{equation}\label{eqn:44}
\frac{\sqrt{3}}{\sqrt{1-\eta}}K\big(\eta/(1-\eta)\big)=R\sqrt{|a|}~,
\end{equation}
where $K(x)=(\pi/2)[1+x/4+(9/64)x^2+(25/256)x^3+...]$ is the complete elliptic integral of the first kind, and $\eta=bm^2(0)/2|a|\le1/2$. Note that $\eta \le 1/2$, just implies the physical requirement that $m(0)^2 \le m_0^2 = |a|/b$. Based on the approximations above, the mean-field force in Eq.~(\ref{eqn:42}) can be recast in terms of $\eta$ as 
\begin{eqnarray}\label{eqn:45}
F_R^{(\text{mf})}=\frac{2\pi\hbar^2N_0LR}{3M}\frac{\eta|a|^2}{b}(1-\eta),~~
\end{eqnarray}
where $\eta$ is to be obtained from the graphical solution of Eq.~\eqref{eqn:44}, which does not appear in closed form. However, an approximate closed form of $\eta$ was obtained by Biswas \textit{et al} in the context of film geometry \cite{Biswas}. Following their approach, for cylindrical geometry, we get
\begin{equation}\label{eqn:46}
\eta(x)=\frac{1}{2}\text{tanh}^2\bigg(\frac{1}{\sqrt{6}}\sqrt{x^2-\frac{3\pi^2}{4}}\bigg)~, 
\end{equation}
where $x = R \sqrt{|a|}$. We emphasis that Eq.~\eqref{eqn:46} gives a very close approximation \footnote{For a comparison see-- Fig. 1 in Ref.~\cite{Biswas}} 
of the actual roots of Eq.~\eqref{eqn:44}. It is clear from Eq.~(\ref{eqn:46}) that $\eta\rightarrow1/2$ as $R\rightarrow\infty$, and $\eta=0$ for $R\sqrt{|a|}\le\sqrt{3}\pi/2$. Thus we get the mean-field bulk pressure exerted by the condensate on the cylindrical wall as $F_R^{(\text{mf})}/2\pi RL|_{R\rightarrow\infty}=g\bar{n}^2/6$ in comparison to $g\bar{n}^2/2$ for the film geometry \cite{Biswas2010}. Now, it is easy to obtain the  mean-field Casimir force,   $f_R^{(\text{mf})}=F_R^{(\text{mf})}-F_\infty^{(\text{mf})}$,  from Eqs.~\eqref{eqn:45}-\eqref{eqn:46}, and it is given by 
\begin{equation}\label{eqn:47}
f_R^{(\text{mf})}=-\frac{2\pi\hbar^2N_0LR|a|^2}{3Mb}\left[\frac{1}{4}-\eta(1-\eta)\right].
\end{equation}
We can rewrite the mean-field force in Eq.~\eqref{eqn:47} in terms of the dimensionless density ($\tilde{\rho}(R,g\bar{n})=\frac{g\bar{n}2MR^2}{\hbar^2}$) and substituting the original expressions of $a$ ($=\frac{2Mg\bar{n}}{\hbar^2}$), $b$ ($=\frac{2MgN_0}{\hbar^2}$) and the coupling constant g ($=\frac{4\pi\hbar^2a_s}{M}$), as
\begin{eqnarray}\label{eqn:72}
f_R^{(\text{mf})}=-\frac{1}{48}\frac{\hbar^2L}{MR^4}\frac{R}{a_s}\tilde{\rho}^2~\text{sech}^4\bigg(\frac{\sqrt{\frac{4\tilde{\rho}}{3}-\pi^2}}{2\sqrt{2}}\bigg).
\end{eqnarray}

\begin{figure}
\includegraphics[width = \linewidth]{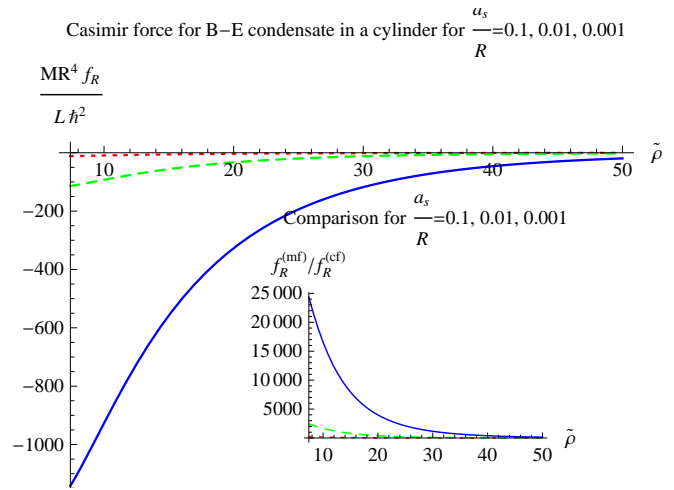}
\caption{(Color online) Casimir force for a Bose-Einstein condensate in a cylinder. The dotted, dashed and blue solid lines follow Eq.~\eqref{eqn:73} for $a_s/R=0.1,~0.01$ and $0.001$ respectively. The inset shows the ratio of the mean-field and the quantum fluctuation parts in same equation for the same interaction strengths ($a_s/R$) with the same line-types.
\label{fig2}}
\end{figure}

\subsection{Total Casimir force }
The  total Casimir force is now simply given by the sum of the terms on the RHS of Eq.~\eqref{eqn:71} and Eq.~\eqref{eqn:72}, as 
\begin{eqnarray}\label{eqn:73}
f_R=f_R^{(\text{mf})}+f_R^{(\text{qf})}~.
\end{eqnarray}
We emphasis that the mean-field part of the Casimir force is a direct consequence of the inhomogeneity of the condensate, and it dominates over the phononic contribution for low densities. For high densities,  the condensate inside the cylinder becomes  essentially homogeneous, and the phononic contribution to the Casimir force dominate  the mean-field part of the Casimir force for typical values of the parameters. We plot this force in units of $\frac{\hbar^2L}{MR^4}$ in Fig.~\ref{fig2}, for different values of interaction strength - $a_s/R$. 

While the Casimir-Polder force between a cigar shaped BEC and a flat surface has been observed to be attractive \cite{Harber}, the Casimir effect for a BEC inside a circular-cylinder although has not been observed; yet is predicted, according to Eq.~\eqref{eqn:73}, to be attractive. This prediction is potentially relevant for atomic waveguides, i.e. geometries in which BECs can be transported. If the velocity of the BEC in the atomic waveguide is less than the Landau critical velocity beyond which the BEC can be destroyed by inelastically interacting with the wall, then the Bogoliubov excitations in the moving condensate would be unaltered from that for the static case. So, the Casimir force for the slowly moving condensate in an atomic waveguide is expected to be attractive along the radial ($\hat{\rho}$) direction, and it may not affect the flow along the perpendicular ($z$) direction. On the other hand, if the velocity of the condensate is more than the Landau critical velocity, then the 
condensate can inelastically interact with the wall. Since the attractive Casimir force between any two parts (semi-cylindrical walls) of the cylindrical tube may induce shrinking of its cross-section, there would be an enhancement of inelastic collisions with the cylindrical wall. Thus, the attractive Casimir force may slowdown the condensate if it is strong enough to shrink the cross-section. This is possible for a sufficiently small radius of a cylindrical waveguide made up of a flexible wall. However, such forces are usually not strong enough to deform cylindrical wall of a metallic waveguide.

\section{Conclusion}
\label{Sec4}

In this article, we have presented theoretical calculation on quantum Casimir force (and pressure) exerted by a self-interacting Bose-Einstein condensate confined to a circular-cylinder by applying zeta regularization technique introduce be Gosdzinsky and Romeo \cite{Gosdzinsky} for the leading order in condensate density, and dimensional regularization cum Chowla-Selberg lattice summation technique \cite{Milton2010} for the higher orders in condensate density. In this regard, we have approximately computed the (grandcanonical) self-energy of the confined condensate up-to the quadratic order in fluctuations. Both the inhomogeneity of the order parameter of the confined BEC and the vacuum fluctuations of the phonon field in the BEC contribute to the Casimir force.  

Our analytic results, for the Casimir force, are valid only for the repulsive interaction ($a_s>0$). For $a_s<0$, the condensate becomes unstable beyond a critical number of particles \cite{Biswas2009}. The Casimir force exerted by the condensate would be minimum for $\tilde{\rho} = 3 \pi^2/4$. For high density, the Bose condensate inside the cylinder essentially becomes homogeneous, and the quantum fluctuation of the phonon field become important. For noninteracting case, there will not be any Bogoliubov excitation. So, fluctuation part of the Casimir force would be zero. But, the mean field part of the Casimir force is not necessarily zero, as because, the profile of the order parameter is still given by $m_{p_z,0,1}(z,\rho,\theta)=\sqrt{\frac{2}{R^2L}}\text{e}^{i\frac{p_zz}{\hbar}}\frac{J_0(\alpha_{0,1}\rho/R)}{J_{1}(\alpha_{0,1})}$\footnote{If the coupling constant ($g$) be zero, there will be no difference between the order parameter and the ground state cavity mode $\phi_{p_z,0,1}(z,\rho,\theta)$.} 
whose derivative at the cylindrical surface ($\rho=R$) is nonzero. Our present calculation can be extended for finite temperature by further consideration of the confinement of thermal fluctuations of the thermal cloud over the condensate \cite{Biswas2007}.

In a realistic experimental scenario, liquid $^4$He can be placed inside a cylindrical shell. Well below the $\lambda$-point ($T_\lambda=2.18$ K), the liquid can become a Bose-Einstein condensate, and the condensate would exert Casimir force on the cylindrical surface. For 1 $\mu$m radius of the shallow of the ultra-cold $^{4}$He liquid ($a_s=2.5\text{\AA}$ \cite{Leggett}), where there is no classical critical fluctuations, the Casimir pressure is predicted from Eq.~\eqref{eqn:73} to be of the order of $-20\times10^{-12}$ N/m$^2$ for the unit dimensionless density ($\tilde{\rho}\rightarrow1$).

While McMahon asymptotic expansion of roots of the Bessel functions and their first derivative fails to reproduce the fluctuation part of the Casimir force as one can obtain from the prediction of Milton, DeRaad Jr., Gosdzinsky and Romeo \cite{Milton1981,Gosdzinsky}, analytic continuation of $\sum_{n=1}^\infty\alpha^{-2}_{\sigma,n}$ from the exact result ($1/4(\sigma+1)$ \cite{Sneddon1960}) may open a door to the exact evaluation of $\sum_{\sigma,n}\alpha^{2}_{\sigma,n}$ and to reproduce the Dirichlet b.c. part. This method is to be generalized for the Neumann b.c. part to predict the actual scenario from the roots of the first derivative of the Bessel functions of the first kind.

Casimir effect could have been further studied by considering a small gap between the condensate and the cylindrical surface generalizing the effect for two concentric cylinders \cite{Teo}. Instead of keeping the BEC inside the cylinder, we could keep a Fermi liquid (say $^3$He). How to calculate the Casimir force, for a Fermi liquid in the confined geometry, is an open question.

\acknowledgments 
S. Biswas and A.A. gratefully acknowledge funding from the INSPIRE Faculty Award by the DST  (Govt. of India). S. Biswas further acknowledges the hospitality of the Department of Physics, IIT-Kanpur during the initial phase of this work. A.A. also acknowledges funding from the Faculty Initiation Grant of IIT-Kanpur, India. Several useful discussions with J.K. Bhattacharjee (HRI, India) and S. Dutta Gupta (University of Hyderabad, India) are gratefully acknowledged. Useful comments of K.A. Milton (University of Oklahoma, USA) prior to the preparation of the present form of the manuscript is also gratefully acknowledged.

\appendix
\section{Chowla-Selberg Lattice Summation}
\label{A}
Chowla-Selberg lattice summation formula involves evaluation of Epstein zeta function \cite{Chowla-Selberg}
\begin{eqnarray}\label{eqn:ap1}
 Z(s)=\Sigma^{'} (am^2+bmn+cn^2)^{-s},
\end{eqnarray}
where $s$ is a complex number, and the summation is for all integers $m,n$ (each going from $-\infty$ to $+\infty$), while the prime indicates that $m=n=0$ is excluded from the summation; further $a>0$ and $c>0$, while $b$ is real number and subject to the condition that the `determinant' $\triangle=4ac-b^2>0$. This is called lattice summation because the summation is over the lattice points ($\{m,n\}$) on the two-dimensional $m-n$ plane. The Epstein zeta function $Z(s)$ expressed in terms of the lattice summation over $m$ and $n$ in Eq.~\eqref{eqn:ap1}, is defined for $\Re[s]>1$, and can be continued analytically over the whole s-plane, and satisfies a functional equation similar to the one satisfied by the Riemann zeta Function \cite{Chowla-Selberg}. Such an analytic continuation, in the context of Casimir effect on different cylindrical geometries, was greatly exercised by Abalo-Milton-Kaplan in Ref.~\cite{Milton2010}. For the evaluation of the Epstein zeta function, we have Chowla-Selberg lattice 
summation formula, as \cite{Chowla-Selberg,Milton2010}
\begin{eqnarray}\label{eqn:ap2}
Z(s)&=&\Sigma^{'} (am^2+bmn+cn^2)^{-s}\nonumber\\&=&2a^{-s}\zeta(2s)+\frac{2^{2s}\sqrt{\pi}a^{s-1}}{\Gamma(s)\triangle^{s-1/2}}\zeta(2s-1)\zeta(s-1/2)\nonumber\\&&+\frac{2^{s+3/2}\pi^s}{\Gamma(s)\triangle^{s/2-1/4}\sqrt{a}}\sum_{n=1}^\infty n^{s-1/2}\sigma_{1-2s}(n)\nonumber\\&&\times\cos(n\pi b/a)2K_{1/2-s}(n\pi\sqrt{\triangle}/a)~, 
\end{eqnarray}
where $K_{\nu}$ is the modified Bessel function of the second kind of order $\nu$, and $\sigma_{k}(n)$ is a divisor function, which is defined as sum of $k$-th powers of the divisors of $n$, as
\begin{eqnarray}\label{eqn:ap3}
\sigma_k(n)=\sum_{d|n}d^k~;
\end{eqnarray}
e.g. divisors of $n=6$ are $1,2,3,6$, so that $\sigma_0(6)=1^0+2^0+3^0+6^0=4$, $\sigma_1(6)=1^1+2^1+3^1+6^1=12$, $\sigma_2(6)=1^2+2^2+3^2+6^2=50$, etc.

For a square-cylinder of cross-sectional area $\tilde{L}^2=\pi R^2$, which is equal to that of the circular-cylinder of radius $R$, $\alpha_{\sigma,n}^2$ in Eq.~\eqref{eqn:13o1} would be $\alpha_{\sigma,n}^2=(\sigma^2+n^2)\pi^2$ where $\sigma=1,2,3,...$ and $n=1,2,3,...$. The space of the lattice summation over $\sigma$ and $n$ is now one fourth of that over $\sigma$ and $n$ in Eq.~\eqref{eqn:ap2} except the summation over the axes ($\sigma=0$ \& $n\neq0$ and $\sigma\neq0$ \& $n=0$) on the $\sigma-n$ plane. Thus we recast Eq.~\eqref{eqn:13o1}, for $s\rightarrow-1$, as
\begin{eqnarray}\label{eqn:ap4}
D^{(-2s)}&=&\frac{1}{4\sqrt{\pi}}\sum_{\sigma=1,n=1}^{\infty,\infty}\left(\frac{\alpha^2_{\sigma,n}}{\tilde{L}^2}\right)^{-s}\frac{\Gamma(s)}{\Gamma(\frac{1+2s}{2})}\nonumber\\&=&\frac{1}{4\sqrt{\pi}}\frac{\Gamma(s)}{\Gamma(\frac{1+2s}{2})}\bigg(\frac{1}{\pi R^2}\bigg)^{-s}\sum_{\sigma=1,n=1}^{\infty,\infty}\left(\alpha^2_{\sigma,n}\right)^{-s}\nonumber\\&=&\frac{\pi^2}{4\sqrt{\pi}}\frac{\Gamma(s)}{\Gamma(\frac{1+2s}{2})}\frac{1}{(\pi R^2)^{-s}}\frac{1}{4}\bigg[\Sigma'\left(\sigma^2+n^2\right)^{-s}\nonumber\\&&-4\zeta(2s)\bigg].~~~~~~
\end{eqnarray}
Analytic continuation of Eq.~\eqref{eqn:ap4} can be obtained by using the lattice summation formula in Eq.~\eqref{eqn:ap2} and the reflection property of the zeta function, 
\begin{eqnarray}\label{eqn:ap5}
\Gamma(s)\zeta(2s)=\Gamma\big(\frac{1-2s}{2}\big)\zeta(1-2s)\pi^{2s-1/2},
\end{eqnarray}
for $s\rightarrow-1$, as \cite{Milton2010}
\begin{eqnarray}\label{eqn:ap6}
D^{(2)}=0.00483155/\pi R^2=0.00153793/R^2.
\end{eqnarray}

For Neumann boundary condition on the square-cylinder, on the other hand, we have $\beta_{\sigma,n}^2=\alpha_{\sigma,n}^2$ except this time $\sigma=0,1,2,...$ and $n=0,1,2,...$. So, this times, half of the two axes of $\sigma-n$ plane would contribute to the lattice summation unlike that in Eq.~\eqref{eqn:ap4}. This would result the only difference between $N^{(2)}$ and $D^{(2)}$. Thus $N^{(2)}$ in Eq.~\eqref{eqn:13o2}, for the square-cylinder, would be \cite{Milton2010}
\begin{eqnarray}\label{eqn:ap7}
N^{(2)}=D^{(2)}-\frac{\zeta(3)}{8\pi^2R^2}=-\frac{0.0429968}{\pi R^2}=-\frac{0.0136863}{R^2}.~~~
\end{eqnarray}

Eq.~\eqref{eqn:ap4} obtained for the square-cylinder of cross-sectional area $\pi R^2$ is not only true for $s\rightarrow-1$ but also for any $\Re[s]\le0$ as far as dimensional regularization and subsequent analytic continuation is concerned. Thus, following the similar steps, as described from Eq.~\eqref{eqn:ap4} to Eq.~\eqref{eqn:ap7}, we get $N^{(4)}=0.00914223/R^4$, $D^{(4)}=-0.000263472/R^4$, $N^{(6)}=-0.0185855/R^6$, $D^{(6)}=-0.000389195/R^6$, $N^{(8)}=0.0745135/R^8$, $D^{(8)}=0.00400674/R^8$, etc. On the other hand, for negative half-integral values of $s$, since $\Gamma(\frac{1+2s}{2})\rightarrow\infty$ in Eq.~\eqref{eqn:ap4}, we have $D^{(-2s)}=N^{(-2s)}=0$.

\begin{thebibliography}{99}
\bibitem{Casimir}H. B. G. Casimir, Proc. K. Ned. Akad. W. \textbf{51}, 193 (1948).
\bibitem{Krech}M. Krech and S. Dietrich, \href{http://dx.doi.org/10.1103/PhysRevA.46.1886}{\pra{46}{1886}{1992}}.
\bibitem{Golestanian}M. Kardar and R. Golestanian, \href{http://dx.doi.org/10.1103/RevModPhys.71.1233}{\rmp{71}{1233}{1999}}.
\bibitem{Gambassi}A. Gambassi, 
\href{http://dx.doi.org/10.1088/1742-6596/161/1/012037}{J. Phys.: Conf. Ser. \textbf{161}, 012037 (2009)}.
\bibitem{Mohideen}U. Mohideen and A. Roy, \href{http://dx.doi.org/10.1103/PhysRevLett.81.4549}{\prl{81}{4549}{1998}}; G. Bressi, G. Carugno, R. Onofrio and G. Ruoso, \href{http://dx.doi.org/10.1103/PhysRevLett.88.041804}{\prl{88}{041804}{2002}}. 
\bibitem{Chan2}A. Ganshin, S. Scheidemantel, R. Garcia, and M. H. W. Chan, \href{http://dx.doi.org/10.1103/PhysRevLett.97.075301}{\prl{97}{075301}{2006}}.
\bibitem{Hertlein}C. Hertlein, L. Helden, A. Gambassi, S. Dietrich and C. Bechinger, \href{http://dx.doi.org/10.1038/nature06443}{\nat{451}{172}{2008}}.
\bibitem{Emig}T. Emig, A. Hanke, R. Golestanian, and M. Kardar, \href{http://dx.doi.org/10.1103/PhysRevLett.87.260402}{\prl{87}{260402}{2001}}.
\bibitem{Chen}F. Chen, U. Mohideen,  G. L. Klimchitskaya, and V. M. Mostepanenko, \href{http://dx.doi.org/10.1103/PhysRevLett.88.101801}{\prl{88}{101801}{2002}}. 
\bibitem{Dalvit}D. A. R. Dalvit, F. C. Lombardo, F. D. Mazzitelli, and R. Onofrio, \href{http://dx.doi.org/10.1103/PhysRevA.74.020101}{\pra{74}{020101(R)}{2006}}.
\bibitem{Lambrecht}A. Lambrecht and V. N. Marachevsky, \href{http://iopscience.iop.org/1742-6596/161/1/012014}{J. Phys.: Conf. Ser. \textbf{161}, 012014 (2009)}; F. Impens, A. M. C. Reyes, P. A. Maia Neto \textit{et al}, \href{http://dx.doi:10.1209/0295-5075/92/40010}{\epl{92}{40010}{2010}}.
\bibitem{Graham}N. Graham, A. Shpunt, T. Emig, S. J. Rahi, R. L. Jaffe, and M. Kardar, \href{http://dx.doi.org/10.1103/PhysRevD.83.125007}{\prd{83}{125007}{2011}}.
\bibitem{Teo}L. P. Teo, \href{http://dx.doi:10.1209/0295-5075/96/10006}{\epl{96}{10006}{2011}}.
\bibitem{Biswas2010}S. Biswas, J. K. Bhattacharjee, D. Majumder \textit{et al}, \href{http://dx.doi.org/10.1088/0953-4075/43/8/085305}{\jpb{43}{085305}{2010}}.
\bibitem{Biswas}S. Biswas, J. K. Bhattacharjee, H. S. Samanta, S. Bhattacharyya, and B. Hu, \href{http://dx.doi.org/10.1088/1367-2630/12/6/063039}{\njp{12}{063039}{2010}}.
\bibitem{Dohm2014}V. Dohm, \href{http://dx.doi.org/10.1103/PhysRevE.90.030101}{\pre{90}{030101(R)}{2014}}.
\bibitem{Diehl2014}H. W. Diehl \textit{et al}, \href{http://dx.doi.org/10.1103/PhysRevE.89.062123}{\pre{89}{062123}{2014}}.
\bibitem{chip}J. Zou, Z. Marcet \textit{et al}, \href{http://dx.doi.org/10.1038/ncomms2842}{Nat. Commun. \textbf{4}, 1845 (2013)}.
\bibitem{Milton1981}L. L. DeRaad Jr. and K. A. Milton, \href{http://dx.doi.org/10.1016/0003-4916(81)90097-X}{Ann. Phys. \textbf{136}, 229 (1981)}.
\bibitem{Gosdzinsky}P. Gosdzinsky and A. Romeo, \href{http://dx.doi.org/10.1016/S0370-2693(98)01164-2}{\plb{441}{265}{1998}}. 
\bibitem{Milton1999}K. A. Milton, A. V. Nesterenko, and V. V. Nesterenko, \href{http://dx.doi.org/10.1103/PhysRevD.59.105009}{\prd{59}{105009}{1999}}.
\bibitem{Milton2010} E. K. Abalo, K. A. Milton, and L. Kaplan, \href{http://dx.doi.org/10.1103/PhysRevD.82.125007}{\prd{82}{125007}{2010}}. 
\bibitem{Straley}J. P. Straley, G. A. White, and E. B. Kolomeisky, \href{http://dx.doi.org/10.1103/PhysRevA.87.022503}{\pra{87}{022503}{2013}}.
\bibitem{Milton2004}For a review, see -- K. A. Milton, \href{http://dx.doi.org/10.1088/0305-4470/37/38/R01}{\jpamg{37}{R209}{2004}}.
\bibitem{Jackson}J. D. Jackson, \textit{Classical Electrodynamics}, 3rd Ed., Sec. 8.2 \& 8.6 (New York: John Wiley, 1999).
\bibitem{Chowla-Selberg}S. Chowla and A. Selberg, \href{http://dx.doi.org/10.1073/pnas.35.7.371}{Proc. Natn'l. Acad. Sc. (USA) \textbf{35}, 371 (1949)}.
\bibitem{McMahon}For higher orders of the asymptotic expansion, see -- J. McMahon, \href{http://dx.doi.org/10.2307/1967501}{Ann. Math. \textbf{9}, 23 (1895)}.
\bibitem{Cavity-Modes}For the cylindrical cavity modes ($\phi_{p_z,\sigma,n}(z,\rho,\theta)$) of the scalar field inside the circular-cylinder, specially for Dirichlet boundary condition, see -- S. Bhattacharyya and J. K. Bhattacharjee, \href{http://dx.doi.org/10.1103/PhysRevB.60.R746}{\prb{60}{R746}{1999}}. For Neumann boundary condition, Bessel zeros ($\alpha_{\sigma,n}$) in $\phi_{p_z,\sigma,n}(z,\rho,\theta)$ are to be replaced by the zeros of the first derivative ($\beta_{\sigma,n}$) of the Bessel functions of the first kind. See -- Refs.~\cite{Jackson,Gosdzinsky} for similar replacement specially for electromagnetic field inside a conducting circular-cylinder. 
\bibitem{Roberts}D. C. Roberts and Y. Pomeau, \href{http://dx.doi.org/10.1103/PhysRevLett.95.145303}{\prl{95}{145303}{2005}}; \href{http://arxiv.org/abs/cond-mat/0503757}{arXiv:cond-mat/0503757}. 
\bibitem{Lee}T. D. Lee, K. Huang, and C. N. Yang, \href{http://dx.doi.org/10.1103/PhysRev.106.1135}{\prv{106}{1135}{1957}}. 
\bibitem{Chan}R. Garcia and M. H. W. Chan, \href{http://dx.doi.org/10.1103/PhysRevLett.83.1187}{\prl{83}{1187}{1999}}.
\bibitem{Harber}D. M. Harber, J. M. Obrecht, J. M. McGuirk, and E. A. Cornell, \href{http://dx.doi.org/10.1103/PhysRevA.72.033610}{\pra{72}{033610}{2005}}.
\bibitem{Biswas2009}S. Biswas, \href{http://dx.doi.org/10.1140/epjd/e2009-00221-7}{Eur. Phys. J. D {\bf 55}, 653 (2009)}. 
\bibitem{Biswas2007}S. Biswas, \href{http://dx.doi.org/10.1088/1751-8113/40/33/002}{\jpamt{40}{9969}{2007}}; S. Biswas, \href{http://link.springer.com/article/10.1140\%2Fepjd\%2Fe2007-00007-y}{\epjd{42}{109}{2007}}.
\bibitem{Leggett}A. J. Leggett, \href{http://dx.doi.org/10.1103/RevModPhys.73.307}{\rmp{73}{307}{2001}}.
\bibitem{Sneddon1960}I. N. Sneddon, Proc. Glasgow Math. Assoc. \textbf{4}, 144 (1960).
\end{thebibliography}
\end{document}